\documentstyle[prb,eqsecnum,aps,epsf,twocolumn]{revtex}

\newcommand{\be}{\begin{equation}}
\newcommand{\ee}{\end{equation}}
\newcommand{\bea}{\begin{eqnarray}}
\newcommand{\eea}{\end{eqnarray}}
\newcommand{\ba}[1]{\begin{array}{#1}}
\newcommand{\ea}{\end{array}}

\begin{document}
\draft
\twocolumn[\hsize\textwidth\columnwidth\hsize\csname @twocolumnfalse\endcsname
 
\title{Coulomb Blockade in a Quantum Dot Coupled Strongly to a Lead}
 
\author{Hangmo Yi and C.\ L.\ Kane}
 
\address{
Department of Physics and Astronomy, University of Pennsylvania,
Philadelphia, PA 19104
}
 
\date{September 21, 1995}
\maketitle
 
{\tightenlines
\begin{abstract}
We study theoretically a quantum dot in the quantum Hall regime
that is strongly coupled to a single lead via a point contact.
We find that even when the transmission through the point contact
is perfect, important features of the Coulomb blockade persist.
In particular, the tunneling into the dot via a second weakly
coupled lead is suppressed, and shows features which can be
ascribed to elastic or inelastic cotunneling through the dot.
When there is weak backscattering at the point contact,
both the tunneling conductance and the differential capacitance are
predicted to oscillate as a function of gate voltage.  We point out
that the dimensionless ratio $\xi$ between the fractional oscillations
in $G$ and $C$ is an intrinsic property of the dot,
which, in principle, can be measured.  We compute $\xi$ within
two models of electron-electron interactions.
In addition, we discuss the role of additional channels.
\\ \\ \\

\end{abstract}

}

]

\narrowtext

\section{Introduction}

The Coulomb blockade occurs in an isolated mesoscopic island
when the capacitive charging energy to add a single electron
suppresses the discrete fluctuations in the island's charge.  \cite{grab}
In the past several years this physics has been
studied extensively in both metallic systems \cite{aver} and in
semiconductor structures. \cite{scot} The Coulomb blockade is most
easily probed by measuring transport through an island
which is weakly coupled to two leads.  By varying a gate
voltage $V_G$, which controls the chemical potential of the island,
peaks in the conductance are observed each time an additional
electron is added.  Between the peaks, the conductance is
activated, reflecting the Coulomb barrier to changing the
number of electrons on the dot.

The coupling to the island via tunnel junctions introduces
fluctuations on the dot which relax the discreteness of
its charge.  When the conductance $\sigma$ of the junctions
is very small, $\sigma \ll e^2/h$, this effect is weak
and gives rise to the well known cotunneling effect,
in which electrons may tunnel virtually through the Coulomb barrier.
\cite{aver2,glat}
Provided one is not too close to a degeneracy between
different charge states, this physics may be described satisfactorily
within low order perturbation theory in the tunnel coupling.

For stronger coupling to the leads, the perturbative analysis
is no longer adequate, and a quantitative description of
the problem becomes much more difficult.  Based on general
arguments, the Coulomb blockade is expected to be suppressed
when $\sigma \approx e^2/h$, \cite{falc} since in that regime, the $RC$ decay
time of the island leads to an uncertainty in the energy which
exceeds the Coulomb barrier.  However, this argument
is unable to predict quantitatively the nature of the suppression
of the Coulomb blockade.

A particularly well suited system to study this physics
is a semiconductor quantum dot at high magnetic fields.
In the integer quantum Hall effect regime, the states
near the Fermi energy of a quantum dot are edge states,
and have a simple, well organized structure, which
is insensitive to the complicating effects of impurities
and chaotic electron trajectories.      In this paper
we shall study a quantum dot in the integer quantum
Hall regime which is strongly coupled to a lead via a quantum
point contact with one (or a few) nearly perfectly transmitting
channels.    

In a recent paper, Matveev \cite{matv} has considered a dot connected
to a single lead by a nearly perfectly transmitting point
contact.  When the transmission of the point contact is
perfect (zero backscattering), he showed that even in the
presence of a substantial Coulomb energy $U$, the differential
capacitance $C = dQ/dV_G$ is independent of $V_G$.  Thus,
the equilibrium charge $Q$ on the island is maximally
un-quantized, since charge is added continuously as the gate
voltage is changed.   He then showed that the presence of
weak backscattering at the point contact leads to weak
oscillations in the differential capacitance as a function
of gate voltage with a period corresponding to the addition
of a single electron.  These oscillations signal the onset
of the quantization of the equilibrium charge on the island.

In addition to the capacitance measurements, \cite{matv,asho}
the fate of the Coulomb blockade in an island
connected to a single lead via a point contact
can be probed by transport, provided an additional lead
is present.  
This leads us to the interesting possibility of 
{\it simultaneously} measuring the conductance
(a transport property) and the capacitance (an
equilibrium property) of a quantum dot.
In this paper, we build on Matveev's work and
compute, in addition to the differential capacitance,
the tunnel conductance and $I$-$V$ characteristics
for an island connected to one lead with
a nearly perfectly transmitting point contact and
connected very weakly to another lead.

\begin{figure}
\epsfxsize=3.1in
\epsffile{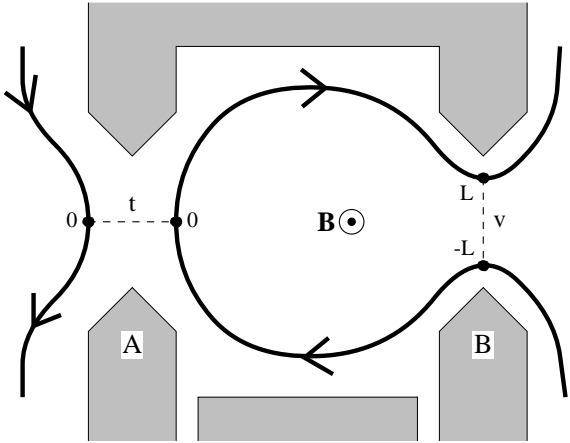}
\caption{Schematic view of a quantum dot connected to leads on both sides.
Negative voltage on the gates (shaded area) confines electrons in the dot.
Thick solid lines show the edge channels formed by strong magnetic field
$B$, where the direction of the electron motion is depicted by arrows.
Dashed lines show tunneling paths, where the tunneling amplitude $t$
and the backscattering amplitude $v$ are controlled by voltage on gate
A and gate B, respectively.}
\label{fig:model}
\end{figure}

Specifically, we consider the quantum
dot depicted schematically in Fig.~\ref{fig:model}.
The center region surrounded by the gates forms a quantum dot,
which is connected to the leads on both sides.
In the $\nu = 1$ quantum Hall regime,
there is a single edge channel going around the dot.
The contact to the left lead
is a tunnel junction, characterized by a small tunneling
amplitude $t$, which is controlled by the voltage on gate A.
The right contact is a nearly perfectly transmitting point
contact characterized by a small backscattering amplitude $v$,
which is controlled by the voltage on gate B.
The backscattering at the point contact
involves tunneling of electrons
between the opposite moving edge channels at $x=L$
and $x= -L$, where $x$ is a parameter specifying
the spatial coordinate along the edge channel.

\vspace{.1in}

A crude estimation of the Coulomb energy with the above
geometry gives $U=K e^2/\epsilon L$, where $K$ is a
dimensionless geometrical factor and $\epsilon$ is the dielectric
constant of the semiconductor material.
Another geometry-dependent energy scale is the level spacing
between edge states in an isolated dot, $\Delta E=\pi\hbar v_F/L$.
While there are few reliable estimates of the edge state
velocity $v_F$, the ratio $\Delta E/U$ is typically of the order 0.1
in GaAs/AlGaAs heterostructures in the integer quantum Hall
effect regime. \cite{vaar}  The smallness of this ratio is assumed
throughout this paper.

Though the equilibrium charge on the dot is not quantized when
the transmission at the point contact is perfect,
we nonetheless find that important features of the
Coulomb blockade remain in the tunneling characteristics.
Remarkably, most of these features can be explained within
the usual {\em weak coupling} model.  
Specifically, we find that at zero temperature, the Ohmic 
conductance is suppressed by a factor of $(\Delta E /U )^2$
below its noninteracting (i.e. $U=0$) value.
This suppression of the tunnel conductance is precisely of
the form predicted by the theory of elastic cotunneling 
through a one dimensional system.  
Evidently, the cotunneling theory is more general than
its derivation within weak coupling perturbation theory suggests.
In addition, when the
temperature $T$ or the voltage bias $eV$ exceeds $\Delta E$,
we find behavior analogous to {\it inelastic} cotunneling.
In particular, for $\Delta E \ll T \ll U$ the tunneling
conductance is suppressed by $(T/U)^2$.  Moreover,
for $\Delta E \ll eV \ll U$, the tunneling current varies as $V^3$.

Inelastic cotunneling behavior has also recently been discussed by
Furusaki and Matveev \cite{furu} for a quantum dot
strongly coupled to two leads.  While similar in many
respects, it is worthwhile to distinguish the present work from
that in Ref.~[\ref{ref:furu}].  In that reference, the two point contacts
are considered to be independent.  An electron passing through
one contact can never coherently propagate to the second
contact.  Elastic cotunneling can therefore not be described within
that framework.  In general, the elastic cotunneling will depend on
the complicated transmission matrix of the sample.  In contrast,
in the quantum Hall regime, the edge channels have a particularly
simple structure and directly connect the leads.  If the phase
coherence length can exceed the dot's dimensions, then elastic
cotunneling should be present.  This physics is correctly accounted
for in our edge state model.  In addition, the beautiful results
in Ref.~[\ref{ref:furu}] concerning the Coulomb blockade for spin
degenerate systems is not applicable here, since the magnetic field
destroys that degeneracy.

In addition to the limiting behaviors discussed above, we
find that at low temperature $T \ll \Delta E$, there
can be nontrivial structure in the $I$-$V$ characteristic
for $V \approx \Delta E$.   
In particular, we find steps in the {\it differential}
conductance, $dI/dV$ as a function of bias voltage.  In the weak
coupling limit, the existence of such steps has been discussed
by Glattli. \cite{glat} They are a consequence of inelastic cotunneling when
the {\it many body} eigenstates of the quantum dot
are discrete.  Observation of such steps could provide a 
new spectroscopy of the low lying many body states of a quantum dot.
Remarkably, as we shall show in Sec.~\ref{sec:pt}, 
these steps can remain sharp in the strong coupling limit, 
even though it becomes meaningless in that limit to
speak of discrete single particle states.

When the transmission through the point contact is less than 
perfect,  charge quantization is introduced in the dot.
The rigidity of the quantization grows with increasing
amplitude of the backscattering at the contact.
For weak backscattering, this is reflected in oscillations in 
the differential capacitance as a function of 
gate voltage, with an amplitude proportional
to the backscattering matrix element $v$.
We find that similar oscillations, proportional to $v$, should be present
in the conductance.   
A comparison of these oscillations should therefore provide 
information about the intrinsic structure of the quantum dot.
We focus on the ratio of the fractional oscillations in the
conductance to that of the capacitance,
\be
\xi = {G_1/G_0 \over {C_1/C_0}}.
\ee
$G_0$ and $C_0$ are the average capacitance and conductance,
whereas $G_1$ and $C_1$ are the amplitudes of the oscillations
as a function of gate voltage.  
By considering the fractional oscillations, $G_1/G_0$
and $C_1/C_0$, the dependence on the tunneling matrix element
$t$ necessary to measure $G$ and the capacitive lever arm $\eta$
necessary to measure $C$ is eliminated.
Moreover, since in the weak backscattering limit both quantities
are proportional to $v$, the dependence on $v$ is eliminated
by taking their ratio.  Thus, $\xi$ measures an intrinsic property
of the strongly coupled quantum dot which is independent of the
details of the tunneling matrix elements.
We have computed $\xi$ within a constant interaction model,
in which the Coulomb interaction couples only to the
total number of electrons on the dot.  
We find that $\xi \approx 1.59$ for $T,\Delta E \ll U$, independent
of $T$, $\Delta E$
and $U$.  More generally, $\xi$ will depend on the specific form of
the electron-electron interactions.

With a few modifications, the above considerations can be extended to
the case in which there are more than one well transmitted channels.
We find that the low bias linear conductance is less suppressed as
the number of channels ${\cal N}$ increases.  It has been recently shown that
the suppression factor becomes $(\Delta E/U)^{2/{\cal N}}$ in a constant
interaction model. \cite{flen}  However, if we take it into considerations
that the interaction strength may differ within the same channel and between
different channels, we find that the factor still has the
form $(\Delta E/U)^2$.  In addition, the ratio $\xi$ depends
sensitively on the form of the electron-electron
interactions between different channels on the dot.
It is equal to zero if different channels do not interact and grows
with increasing strength of the inter-channel interactions.
Thus, $\xi$ is a measure of the inter-channel interaction strength.

This paper is organized as follows.  In Sec.~\ref{sec:iv}, we describe
the model and derive the $I$-$V$ characteristic equation.
In Sec.~\ref{sec:pt} we consider a single channel system with a perfectly
transmitting point contact.  We compute the current and discuss various
features of the result.
In Sec.~\ref{sec:ws}, we consider the effect of the weak backscattering
to the conductance in connection with the effect to the capacitance.
We compute $\xi$ within two specific models of electron-electron interactions.
In Sec.~\ref{sec:mc}, we generalize the results of
Sec.~\ref{sec:pt} and Sec.~\ref{sec:ws} to multiple channel systems.
Finally, some concluding remarks are given in Sec.~\ref{sec:conc}.

\section{Edge State Model}
\label{sec:iv}

We begin in this section by describing our edge state model of a quantum
dot strongly coupled to a single lead.
Here we will consider the case in which only
a single channel is coupled to the lead by a nearly perfectly
transmitting point contact.  Later, in section V we will
consider the case where more than one channel is transmitted.

In the absence of interactions, it is a simple matter to describe
this system in terms of the free electron edge state eigenstates.
Such a description is inconvenient, however, for describing
effects associated with the Coulomb blockade, which are due to
the presence of a Coulomb interaction.   The bosonization technique,
however, allows for an exact description of the low energy
physics in the interacting problem. \cite{hald}

At energies small compared to the bulk quantum Hall energy gap,
the many body eigenstates are long wavelength edge magnetoplasmons.
These may be described as fluctuations in the one dimensional edge
density, $n(x)$, where $x$ is a coordinate along the edge.
Following the usual bosonization procedure, we introduce a
field $\phi(x)$, such that $n(x) = \partial_x\phi/2\pi$.
The Hamiltonian, which describes the compressibility of the edge
may then be written,
\be
H_0 = \int dx {v_F\over {4\pi}} (\partial_x\phi)^2. \label{eq:H0}
\ee
The dynamics of the edge excitations follow from the Kac Moody
commutation relations obeyed by $\phi$,
\be
\left[\frac{\partial_x\phi (x)}{2\pi},\phi (x')\right] = i\delta (x-x'). \label{eq:KM}
\ee
Using (\ref{eq:H0}) and (\ref{eq:KM}) it may easily be seen that
$\partial_t n = v_F \partial_x n$, so that the edge excitations
propagate in a single direction at velocity $v_F$ along the edge.
In this language, the electron creation operator on the edge may be written
as $\psi^\dagger (x) = e^{i\phi(x)}$.

Equations (\ref{eq:H0}) and (\ref{eq:KM}) are an exact description
of a single edge
channel of noninteracting electrons.  We may easily incorporate
interactions into this description.  Specifically, we consider
a ``constant interaction model'' in which the Coulomb interaction
couples to the total number of electrons $N$ on the dot, which depends
on the edge charge between $x=-L$ and $x=L$,
\be
N = {1\over {2\pi}}[\phi(L) - \phi(-L)].
\ee
The self capacitance and the coupling to a nearby gate
can then be described by the Hamiltonian,
\be
H_U + H_G = {U\over 2} N^2 + e\eta V_G N, \label{eq:HU}
\ee
where $\eta$ is a ``lever arm'' associated with the
capacitance coupling to the gate.
Since the interaction is still quadratic in the boson fields,
it may be treated exactly in this representation.

Now we consider tunneling between the edge channels.
We consider the left lead in Fig.~\ref{fig:model} to be a Fermi liquid. 
Without loss of generality, we model it as
another $\nu = 1$ quantum Hall edge, characterized
by a boson field $\phi_l$ with a Hamiltonian $H_0$ in (\ref{eq:H0}).
Tunneling from the left lead into the dot at $x=0$
is then described by the operator
\be
{\sf T}_\rightarrow = \mbox{e}^{i[\phi(0)-\phi_l(0)]} \equiv \mbox{e}^{i\theta}.
\ee
where we define $\theta \equiv \phi(0)-\phi_l(0)$.
For the reverse process, ${\sf T}_\leftarrow = \exp -i\theta$.
The left point contact may be characterized by a tunneling
Hamiltonian,
\be
H_t = t\cos \theta
\ee
where $t$ is the tunneling matrix element.

Similarly, tunneling at the right point contact
involves transfer of electrons between $x = -L$ and $x=L$.  
The Hamiltonian describing these processes can be expressed as
\be
H_v = v\cos[\phi(L)-\phi(-L)] = v\cos 2\pi N, \label{eq:Hv}
\ee
where $v$ is the backscattering matrix element.

It is convenient to eliminate the linear gate voltage term
in (\ref{eq:HU}) by the transformation $N \rightarrow N - N_0$,
where $N_0 = e\eta V_G/(\Delta E + U)$ is the optimal number
of electrons on the dot.
Our model Hamiltonian describing a quantum dot with a
single channel coupled to a lead may then be written
\bea
H & = & H_0[\phi_l] + H_0[\phi] + {U\over 2} N^2 \nonumber \\
  & & + t\cos\theta + v \cos 2\pi (N-N_0). \label{eq:Htot}
\eea

We will find it useful in our analysis to represent the partition
function as an imaginary time path integral.
The action corresponding to $H_0$ is then given by
\be
S_0 = \frac{1}{4\pi}\int dxd\tau\: \partial_x \phi
(v_F\partial_x \phi + i\partial_\tau \phi).
\label{eq:S0}
\ee
Since the remaining terms in the Hamiltonian depend only on
$\theta$ and $N$, it is useful to integrate out all of the
other degrees of freedom.  The resulting action, expressed in
terms of $\theta(\tau)$ and $N(\tau)$, is then given by
\bea
S_{tot}  & = & \frac{1}{2} \sum_{i\omega_n}
\left[ \ba{cc} \theta(-\omega_n) & N(-\omega_n) \ea \right]
{\sf G}^{-1}
\left[ \ba{c} \theta(\omega_n) \\ N(\omega_n)\ea \right] \nonumber \\
  & & + \int d\tau \left[ t\cos\theta + v\cos 2\pi(N-N_0) \right]
\label{eq:Stot}
\eea
where $\omega_n$ is a
Matsubara frequency.  ${\sf G}^{-1}$ is the inverse of the Green's function
matrix and can be explicitly expressed as
\be
{\sf G}^{-1} = {1\over T} \left[ \ba{cc}
    \frac{|\omega_n|}{4\pi}(1+\mbox{e}^{-\frac{\pi|\omega_n|}{\Delta E}}) & -\frac{\omega_n}{2} \\
    \frac{\omega_n}{2} & \frac{\pi|\omega_n|}{1-\mbox{e}^{-\frac{\pi|\omega_n|}{\Delta E}}} + U
  \ea \right] , \label{eq:Green}
\ee
where $\Delta E \equiv \pi v_F/L$.

We now briefly develop the framework for our calculation of
the tunneling current.
The $I$-$V$ characteristic (or equivalently the tunneling density of
states) may be computed using the action $S_{tot}$ in (\ref{eq:Stot}).
Working perturbatively in the tunneling matrix element $t$, we
may compute the tunneling current in the presence of a DC
bias $V$ using Fermi's golden rule.
\bea
I & = & \frac{\pi et^2}{2\hbar}
\sum_{m,n} e^{-E_m/kT}
\left[ \left| \left< n|{\sf T}_\rightarrow|m \right> \right|^2
\delta(E_n-E_m-eV) \right. \nonumber \\
  &  & \quad \left. - \left| \left< n|{\sf T}_\leftarrow|m \right> \right|^2
\delta(E_n-E_m+eV) \right], \label{eq:Ifermi}
\eea
where $\left|m\right>$ is an eigen state of the unperturbed
Hamiltonian with energy $E_m$.  Since the sum on $n$
is over a complete set of states, we can re-express the above equation as
\be
I = \frac{\pi et^2}{2h} P(eV)
\label{eq:IV}
\ee
where
\be
P(E) =
\int dt\:\mbox{e}^{iEt}
\left<\left[\mbox{e}^{i\theta(t)},\mbox{e}^{-i\theta(0)}\right]\right>. \label{eq:P}
\ee

In order to compute $P(E)$, it is useful to consider
the imaginary time ordered Green's function, which may be
readily computed using path integral techniques.
\be
{\cal P}(\tau)  \equiv  \left< \mbox{T}_\tau \mbox{e}^{i\theta(\tau)}\mbox{e}^{-i\theta(0)} \right> .
\ee
The real time correlation function may then be
deduced by analytic continuation.  The two terms
in the commutater in (\ref{eq:P}) lead to
\be
P(E) = P^>(E) - P^<(E)
\ee
with
\be
P^{>,<}(E) = \int dt e^{iEt}
{\cal P}(\tau\rightarrow it \pm 0^+) 
\ee

Limiting behavior of the tunneling current may be deduced analytically
from the asymptotic behavior of ${\cal P}(\tau)$.  In particular,
at zero temperature, the Ohmic conductance is proportional to the
coefficient of the $1/\tau^2$ term.  In addition, it is possible
to compute $P(E)$ numerically, as is described in the
following section and in more detail in appendix A.

\section{Point Contact with Perfect Transmission}
\label{sec:pt}

In this section, we will consider the case where the transmission through
the right point contact is perfect.  In this limit,
there is no quantization of the charge on the dot.  Since charge
may flow continuously through the point contact, there is no
preferred integer value for the charge.
This may be seen clearly from the Hamiltonian
in equation (\ref{eq:Htot}), where for $v=0$
the dependence on $N_0$ is absent.

It follows that as the gate voltage is varied that there should be
no oscillations in either the differential capacitance or the
conductance.  However, we will show below that tunneling through
a large barrier onto the dot is still {\em blocked} by the
charging energy.  This blockade is a result of the fact that
the tunneling electron has a discrete charge, which cannot
immediately be screened.

To compute the tunneling current, we evaluate ${\cal P}_0(\tau)$,
where the subscript $0$ indicates that $v=0$.
Since the action (\ref{eq:Stot}) is quadratic, we may write,
\bea
{\cal P}_0(\tau) & = &
\mbox{e}^{-\frac{1}{2} \left< \mbox{T}_\tau \left[ \theta(\tau)-\theta(0) \right]^2 \right>} \label{eq:Ptdef}\\
  & = & \exp \left[ -\sum_{\omega_n}
\left(1-\mbox{e}^{i\omega_n\tau}\right) {\sf G}_{\theta\theta} \right] \label{eq:Ptdef2}
\eea
where ${\sf G}_{\theta\theta}$ is the top left element of the matrix ${\sf G}$
defined in (\ref{eq:Stot}).  This may be rewritten as
\be
{\cal P}_0(\tau) = \left( {\pi\tau_c/\beta \over {\sin \pi\tau/\beta}}\right)^2
\exp \left[-2\pi T \sum_{\omega_n} (1-e^{i\omega_n\tau}) {f(\omega_n)\over|\omega_n|}
\right] \label{eq:Ptau}
\ee
where $\tau_c$ is the short time cutoff (which is of order the 
inverse of the cyclotron frequency) and
\be
f(\omega) =  \frac{U\left(1-\mbox{e}^{-\frac{\pi|\omega|}{\Delta E}}\right)^2}{2\pi|\omega| +
U\left(1-\mbox{e}^{-\frac{2\pi|\omega|}{\Delta E}}\right)}.
\ee

The first term in (\ref{eq:Ptau}) describes the response for noninteracting
electrons, $U=0$.   This gives us a purely
Ohmic tunneling current $I = G_{U=0} V$.
$G_{U=0}$ is related to the transmission probability
of a free electron
through the left barrier, $G_{U=0} = (e^2/h) T_L$, where
$T_L \propto t^2$.

In the presence of interactions, the low bias linear conductance
may be deduced from the long time behavior of ${\cal P}_0(\tau)$.
In the strong interaction limit, $U\gg\Delta E$, we may 
estimate the limiting behavior of the exponential factor in
(\ref{eq:Ptau}) by noting that 
$f(\omega)$ is approximately a constant in each of the
following three limits:
\be
f(\omega) \approx \left\{\ba{ll}
  0 & \mbox{if}\  \omega \ll \Delta E, \\
  1 & \mbox{if}\  \Delta E \ll \omega \ll U, \\
  0 & \mbox{if}\  U \ll \omega . \\
\ea\right. \label{eq:Ptauapp}
\ee
At zero temperature, we thus have, to logarithmic accuracy,
\be
{\cal P}_0(\tau) \approx \left(\tau_c\over\tau\right)^2 \exp -2 
\int_{\Delta E}^U {d\omega\over\omega} \label{eq:Pint}
\ee
in the long time limit ($\tau\gg\Delta E^{-1}$).  It then follows
that the linear conductance is suppressed.  Performing the
integral in (\ref{eq:Ptau}) exactly, we find
\be
G =  c_1 {e^2\over h} T_L \left(\Delta E\over U\right)^2. \label{eq:Gsing}
\ee
with $c_1 \approx 3.11$.
This should be compared with the 
theory of elastic cotunneling, which is derived in the
case of weak tunneling through both barriers, $T_L,T_R\ll 1$. 
When the dot is a one dimensional system (as it is for 
quantum Hall edge states), the result has been shown to 
be  
\be
G \propto {e^2\over h} T_L T_R \left({\Delta E\over U}\right)^2.
\ee
Evidently, the $\Delta E/U$ suppression predicted in (\ref{eq:Gsing}) remains
valid all of the way up to $T_R = 1$.

At finite temperatures or voltages, $\Delta E \ll eV,T \ll U$, the lower limit
of the integral in (\ref{eq:Pint}) is cut off by $T$ and $eV$.  
The resulting tunneling current may thus be obtained by setting
$\Delta E =0$ and written,
\be
I  = c_2 {e^2\over h} T_L {(eV)^2 + 4\pi^2 T^2\over U^2} V, \label{eq:ITV}
\ee
with $c_2 = 2\pi^2\mbox{e}^{-2\mbox{\bf C}}/3 \approx 2.07$
where $\mbox{\bf C}$ is Euler's constant.
This result is, again, exactly in accordance with the theory of 
inelastic cotunneling, setting $T_R = 1$.

An alternative interpretation of this suppression of the 
tunneling current has been
pointed out in Ref.~[\ref{ref:matv2}].  Suppose that an electron tunnels into
the dot.  The dot would
minimize its electrostatic energy by discharging
exactly one electron.  According to Friedel sum rule, the number of
{\em added} electrons, which is $-1$ in this case,
is equal to $\delta/\pi$ where $\delta$ is the scattering
phase shift of the one dimensional channel.  As in Anderson
orthogonality catastrophe, \cite{ande}
the suppression factor in the tunneling rate 
is related to the phase shift by
\be
{dI\over dV} \propto \varepsilon^\gamma \label{eq:Aoc}
\ee
where $\varepsilon=\max(\Delta E, T, eV)$ is the low energy cutoff and
\bea
\gamma & = & 2\left(\delta\over \pi\right)^2 \nonumber \\
   & = & 2(-1)^2 = 2. \label{eq:gamma}
\eea
Combining (\ref{eq:Aoc}) and (\ref{eq:gamma}), we may reproduce
(\ref{eq:Gsing}) and ({\ref{eq:ITV}).  Note that
(\ref{eq:gamma}) differs from the usual orthogonality
exponent, $(\delta/\pi)^2$,
by a factor of two.  This is because we are tunneling into the middle of a
``chiral'' system consisting only of right moving electrons (or equally
the end of a one dimensional normal electron gas.)

Finally, we note that in the high bias limit, $eV \gg U$, we
recover the linear $I$-$V$ characteristic with an offset
characteristic of the Coulomb blockade,
\be
I = {e^2\over h} T_L \left( V -  {U\over{2e}}\right).
\ee
This offset is a consequence of the fact that at short
times, the electron which tunnels can not be effectively
screened by the point contact.

In addition to the limiting behaviors described above, we have
computed the $I$-$V$ characteristic numerically at zero temperature,
as explained in appendix A.
Fig.~\ref{fig:dIdV} shows the differential conductance $dI/dV$.
What is most striking is that there are sharp steps whose sizes
are approximately $\Delta E$.  
For a nearly isolated dot, a similar phenomenon
has been pointed out by Glattli, \cite{glat} and can be understood to be
a consequence of inelastic cotunneling through a dot with a 
discrete energy level spectrum.

\begin{figure}
\epsfxsize=3in
\epsfbox{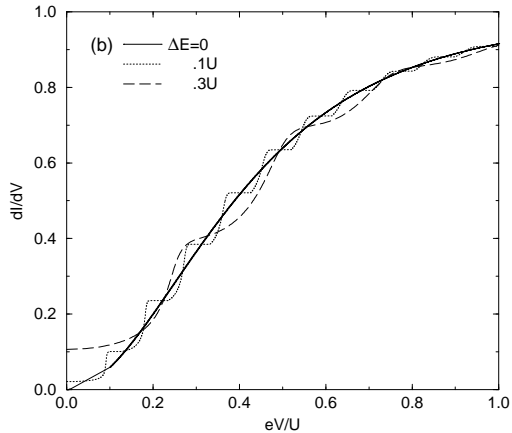}
\caption{Differential conductance $dI/dV$ as a function of scaled bias
voltage $eV/U$, computed in the absence of backscattering ($T_R=1$)
for various $\Delta E$.  Curves are scaled so that $dI/dV=1$ in the
high bias voltage limit.  It is easy to see that the curves
consist of steps of approximate size $\Delta E$.}
\label{fig:dIdV}
\end{figure}
\vspace{.1in}

An inelastic cotunneling process leaves a particle-hole excitation in
the dot.  As the bias voltage increases, the number of available
particle-hole combinations also increases.  Because of
the discrete nature of the energy spectrum of the dot,
this increase in number occurs discontinuously
at every $\Delta E/e$ of the bias voltage, which is manifested
in the tunneling density of states or the differential conductance.
However, the above explanations are not fully adequate in our model because
the dot is strongly coupled to the lead.  For perfect transmission,
the line width of a single particle
energy level is approximately $\Gamma \sim \Delta E$.
It means that the levels are
as broad as the level spacing and the steps are expected to
be wiped out altogether.  The reason for the apparent discrepancy is that
the lifetime of a {\em many body} excited state (i.e.\ a particle-hole pair)
can be much larger than the naive single particle lifetime.

If the dot is weakly coupled to the lead,
the lifetime of a particle-hole excitation may be easily calculated.
First, we
assume the excitation is relaxed only through the process in which both
the particle and the hole tunnel out of the dot.  In analogy with
the theory of cotunneling, \cite{aver2} we use Fermi's golden
rule to estimate the decay rate,
\bea
\tau^{-1} & = & \frac{2\pi}{\hbar}\sum_{k,k'} \left|V_{1k}V_{2k'}\left[\frac{1}{\epsilon_k-(\epsilon_1-\frac{U}{2})}+\frac{1}{\epsilon_2+\frac{U}{2}-\epsilon_{k'}}\right]\right|^2 \nonumber \\
   & & \times \delta (\epsilon_1-\epsilon_2-\epsilon_k+\epsilon_{k'}) \theta(\epsilon_k-\mu)\theta(\mu-\epsilon_{k'}),
\eea
where $\mu$ is the chemical potential of the lead and $\epsilon_1$
and $\epsilon_2$
are the energies of the particle and the hole respectively.  The electron
eigenstates in the lead are labeled by $k$ and $k'$, and $V_{1k}$
and $V_{2k'}$ are matrix elements of the coupling Hamiltonian.
If $\epsilon_1-\epsilon_2 \ll U/2$, the above equation can be approximated
\bea
\tau^{-1} & = & \frac{32\pi}{\hbar} \frac{|V|^4}{\Delta_{lead}^2}\frac{\epsilon_1-\epsilon_2}{U^2} \nonumber \\
  & = & \frac{4}{\pi^2h} (\epsilon_1-\epsilon_2) \left(\Delta E\over U\right)^2 T_R^2, \label{eq:lifetime}
\eea
where $\Delta_{lead}$ is the level spacing of the lead
and $T_R\approx |2\pi V|^2/\Delta E\Delta_{lead}$ is
the transmission probability.
We assume $|V_{1k}|^2 \approx |V_{2k'}|^2 \approx |V|^2$ is constant
in the given range.

In our model,
a simple consideration of the $I$-$V$ characteristic
equation in (\ref{eq:IV}) and (\ref{eq:pt}) shows that the width of the
step risers is proportional to $(\Delta E/U)^2$.  Since the width of the
risers is directly proportional to the line width of the energy levels,
it is evident that Eq.~(\ref{eq:lifetime}) is valid even up to $T_R=1$
with a possible numerical factor.
We thus conclude that even in the presence
of a perfectly transmitting contact,
there can be long-lived excited states in the dot, whose decay is
suppressed by the Coulomb blockade.  

So far we have assumed $\Delta E$ is constant.  If we allow non-uniform
level spacing, degeneracy in the particle-hole excitation energy is
lifted and all steps but the first one split into several
sub-steps.  As the degree of degeneracy increases with the energy,
more splittings occur at higher bias voltages, finally making it
hard to distinguish between steps.

Before we close this section, let us consider other relaxation processes.
It is only when all relaxation rates are less than $\Delta E$ that it is
possible to experimentally observe the steps.  This criterion is equivalent
to saying that the inelastic scattering length, $l_\phi$, is much
longer than the circumference of the dot.  It is known that inelastic
scattering is strongly suppressed in the quantum Hall regime \cite{liu}
most likely due to the difficulty of conserving both energy and momentum
when scattering occurs in a one dimensional channel.  These steps may thus
be observable in the quantum Hall regime.

\section{Point Contact with Weak Backscattering}
\label{sec:ws}
In this section, we consider the case where there is weak backscattering
at the right point contact.
As the contact is pinched off, {\em fractional} charge
fluctuations in the dot are hampered and the discreteness of charge becomes
important.  For weak backscattering $v$, there is an energy cost,
proportional to $v$ in the Hamiltonian (\ref{eq:Hv})
for non-integral charge configurations.
This gives rise to oscillations in physical
quantities such as the capacitance and the conductance as
a function of gate voltage.  The period of these oscillations
corresponds to changing the optimal number of electrons on the
dot $N_0$ by one.

For nearly perfect transmission through the point contact, we thus
expect the conductance and the differential capacitance 
$C = -edN/dV_G$ to have the form,
\bea
C & = & C_0 + C_1 \cos 2\pi N_0 \label{eq:C01} \\
G & = & G_0 + G_1 \cos 2\pi N_0, 
\eea
where the oscillatory components $C_1$ and $G_1$ are proportional
to $v$.  An intrinsic quantity, which is independent of the
backscattering amplitude $v$, the tunneling matrix element $t$,
and the capacitive lever arm $\eta$ associated with the gate, is the
ratio 
\be
\xi \equiv {G_1/G_0 \over {C_1/C_0}}. \label{eq:xi}
\ee

Using the model we have developed so far, we can calculate
the capacitance and the linear conductance perturbatively
in the backscattering matrix element $v$.
The differential capacitance is given by
$C = T d^2 \ln Z / dV_G^2/\eta$, where $Z$ is the partition
function.  As shown by Matveev, to leading order in $v$
the differential capacitance has the form (\ref{eq:C01}) with
average and oscillatory components given by
\bea
C_0 & = & \frac{\eta e^2}{\Delta E+U}, \label{eq:C0} \\
C_1 & = & v \eta\mbox{e}^{-2\pi^2 \left< N(0)^2\right>_0} 
\left( \frac{2\pi e}{\Delta E+U} \right)^2. \label{eq:C1}
\eea
The average $\left<\cdots\right>_0$ is with respect to the ground
state of the unperturbed action $S_{tot}$ in (\ref{eq:Stot}).

In order to compute the 
conductance, we must calculate ${\cal P}(\tau)$ in the presence of the
perturbation $v\cos 2\pi(N - N_0)$.  To the first order in $v$, we find
that ${\cal P}(\tau) = {\cal P}_0(\tau) + {\cal P}_1(\tau) \cos 2\pi N_0$,
where  ${\cal P}_0(\tau)$ is given in (\ref{eq:Ptdef}) and
\bea
{\cal P}_1(\tau) & = & - \left< \mbox{T}_\tau \mbox{e}^{i\left[ \theta(\tau) - \theta(0) \right] } 
S_v \right>_0 \nonumber \\
     & & + \left< \mbox{T}_\tau \mbox{e}^{i\left[ \theta(\tau) - \theta(0) \right] }\right>_0 
\left< S_v \right>_0 
\eea
where $S_v = \int_0^\beta d\tau' v \cos 2\pi N(\tau')$.
This may be written as
\bea
{\cal P}_1(\tau) & = & v{\cal P}_0(\tau) 
\mbox{e}^{-2\pi^2\left<N(0)^2\right>_0} \nonumber \\
      & & \times\int d\tau '\:
\left\{1-\cosh 2\pi\left[{\sf G}_{\theta N}(\tau-\tau')-
{\sf G}_{\theta N}(\tau')\right]\right\}, \nonumber \\
\label{eq:Ptau1}
\eea
where ${\sf G}_{\theta N}(\tau)$ is the off diagonal element of the
Green's function defined in (\ref{eq:Stot}), which may be computed
explicitly using (\ref{eq:Green}).

In order to compute the linear conductance, we must compute
the large $\tau$ limit of (\ref{eq:Ptau1}).  For $U\tau \gg 1$
${\sf G}_{\theta N}(\tau)$ decays as $(U\tau)^{-1}$, so that the integral is independent
of $\tau$.  We thus find
\be
G_1 = v G_0  \mbox{e}^{-2\pi^2\left<N(0)^2\right>_0} 
\times 2 \int d\tau' \left[ 1 - {\rm cosh} 2\pi {\sf G}_{\theta N}(\tau') \right], 
\label{eq:G1}
\ee
where $G_0$ is the zeroth order linear conductance.  

Using (\ref{eq:C0}), (\ref{eq:C1}), and (\ref{eq:G1}),
we obtain an exact expression for the ratio $\xi$,
\be
\xi = { \Delta E + U \over {4 \pi^2}} \times 
2 \int d\tau' \left[ 1 - {\rm cosh} 2\pi {\sf G}_{\theta N}(\tau') \right]. \label{eq:xiG}
\ee
In the limit $T,\Delta E \ll U$, the integral approaches a 
finite value which depends only on $U$.  In this limit
we find
\be
\xi \approx 1.59. \label{eq:xiCI}
\ee

The cancellation of input parameters like $v, U,\ \mbox{and}\ \Delta E$
may tempt us to suspect $\xi$ be a universal number being constant for all
samples.  As will be shown below, however, Eq.~(\ref{eq:xiCI}) is true
only within a constant
interaction model, in which the dependence of the interaction on the
spatial separation is ignored.  In order to see how $\xi$ changes
with different models, let us consider
a more general model whose interaction Hamiltonian is given by
\be
H_U' = \frac{1}{8\pi^2} \int dxdx'\:\partial_x\phi(x)U(x,x')\partial_x\phi(x').
\ee
The constant interaction model is regained by assuming $U(x,x')=U$ to be
uniform.  It is sufficient for our purpose to consider just another
example.  We can think of a local
interaction $U(x,x')=2LU\delta(x-x')\theta(L-|x|)$,
which is certainly an extreme limit
to the other direction from the constant interaction model.
The appropriate Green's function is given by
\be
{\sf G}_{\theta N}(\omega_n) = \frac{T}{\omega_n}\left(1-\mbox{e}^{\frac{\pi|\omega_n|}{U}}\right),
\ee
and then using (\ref{eq:xiG}) we get
\be
\xi = 1.
\ee
Now it is clear that $\xi$ depends on the form
of the electron-electron interaction.
It is, however, noteworthy that the values of $\xi$ computed in two
extreme limits are of the same order of magnitude.

\section{Multiple Channel Systems}
\label{sec:mc}
In this section we generalize the considerations in the previous sections
to the systems of which right contact (nearly) perfectly
transmits more than one channel.  As the conductance of the contact
increases with the number of well transmitted channels ${\cal N}$,
the shorter $RC$ decay time allows a higher uncertainty
in the energy of the island.  Therefore it is natural to expect that
the effect of the Coulomb blockade become weakened in multiple channel
systems, which will be confirmed below.

It turns out that most of the qualitative considerations for the single
channel systems can directly
be applied to multiple channel systems.  Similar calculations as
in Sec.~\ref{sec:pt} show that in the absence of backscattering, the low bias
linear conductance is still suppressed below its noninteracting value,
although the suppression is less strong if ${\cal N}$ is bigger.
On the other hand, there are
new features arising from the introduction of additional channels, on
which we will focus in this section.

For integer quantum Hall states with $\nu >1$, the edge
channels tend to be spatially separated.  The tunneling will
be dominated by the coupling to the nearest edge channel.
As indicated in Fig.~\ref{fig:mc}, this means that
the tunneling and backscattering will occur in different channels.
Then the Hamiltonian for ${\cal N}$ channels is represented by 
\bea
H & = & H_0[\phi_l] + \sum_i H_0[\phi_i] + H_U \nonumber \\
  &   & \quad +\ t\cos\theta_1 + v\cos 2\pi\left(N_{\cal N}-N_0\right),
\label{eq:Htotmc}
\eea
where the notation is similar to that in Sec.~\ref{sec:pt} with the subscript
denoting the channel number except that $\phi_l$ is the boson field
of the left lead.  We have redefined $N_0$
as the optimal number of electrons in channel ${\cal N}$ {\em alone},
which in general depends on lever arms for all channels.

\begin{figure}
\epsfxsize=3.1in
\hskip 0.0truein \epsffile{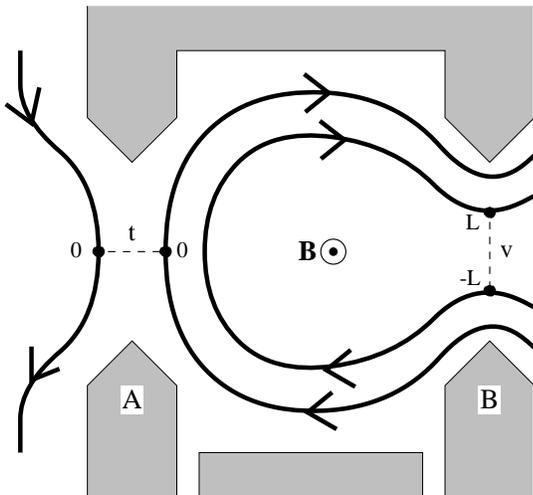}
\caption{Schematic view of a quantum dot analogous to Fig.~1, with two
well transmitted channels.  The symbols are the same as in Fig.~1.
Note that the tunneling through the left contact occurs in the outer
channel whereas the backscattering at the right contact occurs in the inner
channel.}
\label{fig:mc}
\end{figure}
\vspace{.1in}

One of the simplest models to study a multiple channel system is
a constant interaction model in which the interaction Hamiltonian depends
only on the total charge in the island.  The interaction
may be explicitly written
\be
H_U={U\over 2} \left(\sum_i N_i \right)^2.
\ee
The calculation of the tunneling conductance proceeds
along the same lines as in Sec.~\ref{sec:pt}.  In this case, the
function $f(\omega)$ defined in (\ref{eq:Ptau}) is given by
\be
f(\omega) = \frac{U\left(1-\mbox{e}^{-\frac{\pi|\omega|}{\Delta E}}\right)^2}{2\pi|\omega| +
{\cal N} U\left(1-\mbox{e}^{-\frac{2\pi|\omega|}{\Delta E}}\right)}.
\ee
Note that the limiting values of $f(\omega)$ are
\be
f(\omega) \approx \left\{\ba{ll}
  0 & \mbox{if}\  \omega \ll \Delta E, \\
  \frac{1}{\cal N} & \mbox{if}\  \Delta E \ll \omega \ll {\cal N} U, \\
  0 & \mbox{if}\  {\cal N} U \ll \omega . \\
\ea\right. \label{eq:flim}
\ee
It immediately follows that
\be
{dI\over{dV}} \propto \frac{e^2}{h} T_L \left( \frac{\varepsilon}{U} \right)^{2\over{\cal N}}, \label{eq:G2onc}
\ee
where $\varepsilon = \max(\Delta E, eV, T)$.
The exponent $2/{\cal N}$ has been derived in some other papers in several
different contexts. \cite{flen,matv2}
It is clear from (\ref{eq:G2onc}) that the
conductance is less suppressed if there are more channels.

However, it turns out that the nonanalytic behavior with
exponent $2/{\cal N}$ is correct only to the extent that
the constant interaction model is valid.  As explained below,
this is due to a special symmetry of the charging energy
with respect to redistribution of charge among the different channels.
We consider an effective-capacitance model which is one step more general
and has been introduced and developed by several authors to remedy some
problems with the constant interaction model. \cite{kina}
This model assumes that
the edge channels are capacitively coupled metal bodies and the Coulomb
interaction energy depends on the number of electrons in {\em each} channel.
The Coulomb interaction part of the new Hamiltonian can be written
\be
H_U = \frac{1}{2} \sum_{ij} N_i {\sf U}_{ij} N_j, \label{eq:HUmc}
\ee
where ${\sf U}$ is an ${\cal N}\times{\cal N}$ matrix which can be determined
experimentally.  In order to get a clear understanding of 
the effect of this generalization, let us consider a simple specific
example of the electron-electron interaction, i.e.,
\be
{\sf U}_{ij} = u[{\cal N} a\delta_{ij} + (1-a)]. \label{eq:UeffCa}
\ee
The diagonal component $({\cal N} a+1-a)u$ is the magnitude of the
interaction strength within each channel and the off-diagonal
component $(1-a)u$ is that between different channels.
This matrix is chosen such that if the lever arms are all equal to
unity, the total capacitance $C_{tot} \equiv dQ_{tot}/dV_G =
e^2\sum_{ij}{\sf U}_{ij}^{-1} = e^2/u$
in the limit $\Delta E = 0$, independent of $a$.  Note that we regain
a constant interaction model if $a=0$.  As $a$ grows we move away
from the model, finally reaching an independent channel model at $a=1$,
where different channels do not interact.  As in (\ref{eq:Ptauapp}),
the limiting behavior of $f(\omega)$ defined in (\ref{eq:Ptau}) is
given by
\be
f(\omega) \approx \left\{\ba{ll}
  0 & \mbox{if}\  \omega \ll \Delta E, \\
  1 & \mbox{if}\  \Delta E \ll \omega \ll {\cal N} au, \\
  \frac{{\cal N} a+1-a}{\cal N} & \mit {\cal N} au \ll \omega \ll {\cal N} u, \\
  0 & \mbox{if}\  {\cal N} u \ll \omega , \\
\ea\right.
\ee
It is clear from the above equation that if $a\rightarrow 0$ (constant
interaction), (\ref{eq:flim}) is restored and we get the
exponent $2/{\cal N}$, as we discussed earlier.
When $a$ is not small, there is
no appreciable range in which $f(\omega) = 1/{\cal N}$, and the
differential conductance has a different exponent, 2, i.e.
\be
{dI\over{dV}} \propto \frac{e^2}{h} T_L \left(\varepsilon \over u\right)^2
\label{eq:G2}
\ee
provided $\varepsilon=\max(\Delta E, eV, T) \ll {\cal N} au$.

The above considerations can be generalized to a model
with a generic matrix ${\sf U}$.
Even though $f(\omega)$ is a complicated function which depends
on all matrix elements of ${\sf U}$, there exists an energy
scale $\tilde{u}$ which corresponds to $au$ in the above special
model, such that
\be
f(\omega) \approx 1 \quad\mbox{if}\  \Delta E\ll\omega\ll{\cal N}\tilde{u}.
\ee
Then Eq.~(\ref{eq:G2}) is valid if $\varepsilon\ll{\cal N}\tilde{u}$,
except for the factor depending on $a$.  In general, $\tilde{u} = 0$
for a constant interaction model and it measures
how far the used model is away from the constant interaction model.  Van der
Vaart {\em et al.}\cite{vaar2} have measured the
matrix elements of ${\sf U}$ for two Landau
levels confined in a quantum dot.  Even though both contacts were
nearly pinched off in the reference as opposed to our model,
it is suggestive to estimate
the magnitude of $\tilde{u}$ using the experimental data.
With their particular setup,
they got ${\sf U}_{11}=800$, ${\sf U}_{22}=1175$, and ${\sf U}_{12}=650$
(all in $\mu eV$).  A simple estimation with these numbers
gives ${\cal N}\tilde{u} \sim 260 \mu eV \gg \Delta E, T, eV$, which suggests
that (\ref{eq:G2}) must be used rather than (\ref{eq:G2onc}) in this
case.  It has to be admitted that this is a naive estimation considering
the difference between their experimental setup and our theoretical
model.  Opening up a point contact would in general
reduce the strength of electron-electron interactioins in the dot
and it would change the capacitance appreciably.  However, even though
the above estimation of $\tilde{u}$ may be merely speculative at
its best, we expect (\ref{eq:G2}) has to be true if different
channels are weakly coupled, which seems more general
than the constant interaction limit, in practice.

It is now evident that the exponent is $2/{\cal N}$ only for the constant
interaction model.  This is due to a special symmetry of the
constant interaction model, i.e., the interaction part of the Hamiltonian
(\ref{eq:Htotmc}) is invariant under redistribution of total
charge among the different channels.  The effect of the symmetry to the
exponent can be most easily
understood in terms of Anderson orthogonality catastrophe. \cite{matv2}
Eq.~(\ref{eq:Aoc}) can be
directly used with an appropriately generalized definition of $\gamma$, i.e.,
\be
\gamma = 2\sum_i \left(\delta_i\over\pi\right)^2, \label{eq:gammamc}
\ee
where $\delta_i$ is the phase shift in channel $i$.  It needs only a little
consideration of electrostatics to figure out $\delta_i$.  Following
the argument in Sec.~\ref{sec:pt}, let us suppose that an
electron has just tunneled into channel 1 through the left contact.
The number of electrons {\em discharged}
from each channel $-\delta_i/\pi$ depends on the form of the interaction,
provided they satisfy
the constraint $\sum \delta_i/\pi = -1$.  If we work in a constant interaction
model, because the Hamiltonian depends only on the total charge, from the
symmetry $\delta_i/\pi = -1/{\cal N}$ for all $i$.  Therefore
\be
\gamma = 2 \sum_{i=1}^{\cal N} \left(-\frac{1}{\cal N}\right)^2 = \frac{2}{\cal N}.
\ee
On the other hand, if we use an effective-capacitance model and the
system is safely away from the constant interaction
limit (${\cal N}\tilde{u}\gg\Delta E$),
it is always energetically favorable to take a whole electron from
channel 1 and have the exactly same ground state charge configuration as
before.  Then $\delta_1/\pi=-1$ and $\delta_i=0\ (i=2\ldots {\cal N})$,
so that
\be
\gamma = 2 (-1)^2 = 2.
\ee
Then Eqs.\ (\ref{eq:G2}) and (\ref{eq:G2onc})
are readily reproduced from (\ref{eq:Aoc}) and (\ref{eq:gammamc}).
The physical distinction between the energy scales $u$ and $\tilde{u}$
is thus clear.  When an electron is added to the dot, $({\cal N} u)^{-1}$,
which corresponds to the $RC$ decay time, sets
the time scale for the total charge of the dot to return to its original
value.  However, even after the total charge has been screened,
there may be some imbalance in the distribution of charge between
the channels.  $({\cal N}\tilde{u})^{-1}$ sets the scale for the
relaxation of this
imbalance.  In the constant interaction model, there is no
Coulomb energy cost for such an imbalance, so $\tilde{u}\rightarrow 0$.

Now let us consider the effect of weak backscattering.  As in
a single channel model, the introduction of weak
backscattering $v$ results in oscillations in the capacitance and the
conductance.  However, an important difference arises from
the fact that the tunneling and backscattering occur in different
channels.

It has been shown both theoretically \cite{kina} and
experimentally \cite{alph} that the period of the conductance and
the capacitance oscillations
increases with increasing number of well transmitted channels.
This is because the oscillations arise only from the quantization
of $N_{\cal N}$, the number of electrons in the backscattered channel.
When there are many perfectly transmitting channels, many electrons
must be added to the dot to increase $N_{\cal N}$ by 1.

The analysis of the amplitude of the oscillations is
a little more complicated.  We will
again focus on the ratio $\xi$ defined in (\ref{eq:xi}), using the
model interaction in (\ref{eq:UeffCa}).  We
assume $\Delta E, T \ll {\cal N} au$
and all lever arms are taken to be unity.  One may include the
lever arms explicitly, but it does not change the result qualitatively.
Along the same lines as in Sec.~\ref{sec:ws},
the fractional capacitance oscillation may be written
\be
\frac{C_1}{C_0} = v \mbox{e}^{-2\pi^2 \left< N_{\cal N}(0)^2\right>_0} \left(\frac{2\pi}{\cal N} \right)^2 \frac{1}{u}.
\ee
The fractional conductance oscillation may also be written
\be
G_1 = v G_0  \mbox{e}^{-2\pi^2\left<N_{\cal N}(0)^2\right>_0} 
\times 2 \int d\tau' \left[ 1 - {\rm cosh} 2\pi {\sf G}_{\theta N}(\tau') \right], \label{eq:G1mc}
\ee
where the Green's function is given by
\bea
{\sf G}_{\theta N}(\tau) & = & \left<\mbox{T}_\tau \theta_1(\tau)N_{\cal N}(0)\right> \nonumber \\
  & = & -\int d\omega_n\: \mbox{e}^{-i\omega_n\tau} \nonumber \\
  & & \times \frac{(1-a)u\:\mbox{sgn}(\omega_n)}{(2\pi|\omega_n|+{\cal N} au)(2\pi|\omega_n|+{\cal N} u)}.
\eea
We may easily compute ${\sf G}_{\theta N}(\tau)$ in several limits, namely,
\be
{\sf G}_{\theta N}(\tau) \approx \left\{\ba{ll}
  0 & \mbox{if}\  \tau=0, \\
  -i\frac{\pi(1-a)}{\cal N} & \mbox{if}\  \frac{2\pi}{{\cal N} u} \ll \tau \ll \frac{2\pi}{{\cal N} au}, \\
  0 & \mbox{if}\  \tau\rightarrow\infty, \\
\ea\right. \label{eq:Gtn}
\ee
and it is monotonically interpolated in between.  The above equation
is not helpful if $a\sim 1$, but it is sufficient for our purpose which
is to see how $\xi$ changes as the system moves away from the constant
interaction limit.  Since ${\sf G}_{\theta N}(\tau)$
measures the response of $N_{\cal N}$, a period of time $\tau$ after
an electron is added into channel 1, the physical interpretation of the
above limiting behavior is clear.  It takes a time period of
order $2\pi/{\cal N} u$ for the total charge of the dot to return to its
original value.  Channel ${\cal N}$ contributes to this process by
discharging $(1-a)/{\cal N}$ of an electron, which can be read from
the second line of (\ref{eq:Gtn}).  The reason it is proportional
to $1-a$ is that the inter-channel interaction strength is proportional
to $1-a$.  After a time period of order $2\pi/{\cal N} au$, the
charge in {\em each} channel returns to its original value, which
is reflected in the vanishing ${\sf G}_{\theta N}(\tau)$ in the long
time limit.
In order to estimate the integral in (\ref{eq:G1mc}), we may make a
crude approximation by substituting a square function
for ${\sf G}_{\theta N}(\tau)$,
i.e., ${\sf G}_{\theta N}(\tau) = -i\pi(1-a)/{\cal N}$
if $\frac{2\pi}{{\cal N} u} < \tau < \frac{2\pi}{{\cal N} au}$,
and ${\sf G}_{\theta N}(\tau)=0$ otherwise.  Then we get
\be
\frac{G_1}{G_0} \approx v \mbox{e}^{-2\pi^2\left<N_{\cal N}(0)^2\right>_0} \frac{8\pi(1-a)}{{\cal N} au}\sin^2\frac{\pi(1-a)}{2{\cal N}},
\ee
and finally
\be
\xi \approx \frac{2{\cal N}}{\pi}\frac{1-a}{a} \sin^2\frac{\pi(1-a)}{2{\cal N}}.
\ee
This is a good approximation if $a \ll 1$.
This is a monotonically decreasing function of $a$ and as is explained
below, it is a consequence of the fact that the tunneling and the
backscattering occur in different channels.  A bigger $a$
implies weaker inter-channel interactions and consequently a weaker
effect of the backscattering to the conductance.  At $a=1$
(independent channels), we cannot use the above equation,
but we know that the conductance oscillation would eventually
vanish because the backscattering potential does not affect
the conductance at all, and therefore $\xi=0$.
At $a=0$, (constant interaction) $G_1/G_0$ diverges, and so
does $\xi$.  It is because $G_1/G_0$ diverges
as $(u/\varepsilon)^\frac{1}{\cal N}$ if the low energy
cutoff $\varepsilon$ is small,
suggesting that the perturbation theory break down.  Without
detailed calculations, one might have been able to infer it from
the following physical argument.  In a constant interaction model,
the total number of electrons in the dot $\sum_i N_i$ is the only
gapped mode and there are ${\cal N}-1$ combinations of $N_i$
whose fluctuations are not bounded, leading to divergences in
individual terms in the perturbation expansion.  Therefore, we need to sum up
all higher order terms in order to obtain a correct result.  In a series of
recent papers, Matveev and Furusaki \cite{matv,furu} have calculated
both the conductance and the capacitance oscillations nonperturbatively
in a spin-degenerate two-channel model, which they related to the
multichannel Kondo problem.  Their calculations show that the
oscillations are no longer sinusoidal and the period becomes ${\cal N}$
times smaller so that the maximum occurs each time an electron is added
to the dot as a whole (not channel ${\cal N}$ alone).  Such results, however,
clearly apply only in the case where the degeneracy is guaranteed
by a symmetry and hence should not apply in this quantum Hall system.

Without qualitative changes,
the above considerations can be generalized to an effective-capacitance
model with a generic matrix ${\sf U}$.  As in the discussions of
the differential conductance $dI/dV$ earlier in this section,
an energy scale $\tilde{u}$, which plays the role of $au$, can be
determined from the given matrix ${\sf U}$.  In most real situations of
quantum Hall effect edge channels, $\xi$ is a finite quantity
which can be numerically calculated in the effective-capacitance model
if all matrix elements of ${\sf U}$ are known.

\section{Conclusions}
\label{sec:conc}
In this paper we have shown that characteristics of
the Coulomb blockade, which are normally associated with
the weak coupling limit, persist to strong coupling to a lead
via a single channel point contact.  In particular, (i)
we find the analogies of elastic and inelastic cotunneling
in the $(\varepsilon/U)^2$ suppression of the tunnel conductance.
(ii) We find that particle-hole excitations on the dot
can acquire a long lifetime due to a ``Coulomb blockade'' to
relaxation.  This in principle could lead to observable
steps in the low bias differential conductances as a function
of bias voltage.  (iii) The high bias behavior of the $I$-$V$
characteristic has an offset, indicating the presence of a Coulomb gap.
We find similar conclusions when multiple channels are transmitted
through the contact, though the suppression of the Ohmic
conductance is reduced.  In the special case of the constant
interaction model, when there is no penalty towards redistribution
of charge between the channels, the exponent of the suppression is
modified $(\varepsilon/U)^{2\over{\cal N}}$.

When the transmission through the point contact is less then perfect,
the oscillations in the conductance and the capacitance may be
characterized by the dimensionless ratio $\xi$.  While $\xi$ is
independent of the tunneling matrix elements, it depends on the
precise form of the Coulomb interactions.  For a single channel,
we have computed it for two different forms of the
interaction, and its value is of order unity.  For
multiple transmitted channels, its value depends even more sensitively
on the inter-channel interactions, which is zero when
different channels are independent, and grows with increasing
strength of the inter-channel interactions.

\section*{Acknowledgements}
We thank A.\ T.\ Johnson, L.\ I.\ Glazman, T.\ Heinzel, and K.\ A.\ Matveev
for valuable comments and discussions.  This work has been supported
by the National Science Foundation under grant DMR 95-05425.

\appendix

\section{Integral Equation for~$P_0^>(\omega)$}
In this appendix, we will use Minnhagen's integral equation \cite{minn} method
to compute the spectral density
function $P_0(E)$ for the perfect transmission case.  Functions will be
given the subscript $0$ to explicitly show that they are calculated
in the absence of backscattering.  The calculation of the imaginary time Green
function ${\cal P}_0(\tau)$ is straightforward from (\ref{eq:Ptdef2}),
and by analytically continuing it we get
\bea
P_0^>(t) & = & {\cal P}_0(\tau\rightarrow it+0^+) \nonumber \\
       & = & \mbox{e}^{-\left<\theta(0)^2\right>_0} \exp\int_0^\infty d\omega\:\frac{\alpha(\omega)}{\omega}\mbox{e}^{-i\omega t}, \label{eq:pt}
\eea
where the average $\left<\cdots\right>_0$ is evaluated over the unperturbed
action ($t=v=0$).  The function $\alpha(\omega)$ is defined
\bea
\alpha(\omega) & \equiv & \frac{\omega}{2\pi}\int dt\:\mbox{e}^{i\omega t}\left<\theta(t)\theta(0)\right>_0 \nonumber \\
       & = & i\frac{\omega}{2\pi} \left[\left.\left<|\theta(\omega_n)|^2\right>_0\right|_{|\omega_n|\rightarrow i\omega} - \left.\left<|\theta(\omega_n)|^2\right>_0\right|_{|\omega_n|\rightarrow -i\omega}\right] \nonumber \\
       & = & 2 +\frac{2\frac{U}{\pi\omega}\sin\frac{\pi\omega}{\Delta E}\left(1-\cos\frac{\pi\omega}{\Delta E}\right)}{\left(1 + \frac{U}{\pi\omega}\sin\frac{\pi\omega}{\Delta E} \right)^2 - 2\frac{U}{\pi\omega}\sin\frac{\pi\omega}{\Delta E}\left(1-\cos\frac{\pi\omega}{\Delta E}\right)}. \nonumber \\
\label{eq:alpha}
\eea
Now we differentiate (\ref{eq:pt}) with respect to $t$ and Fourier
transform it.  Then we finally get an integral equation
\be
\omega P_0^>(\omega) = \int_0^\omega d\omega '\:\alpha(\omega ')P_0^>(\omega-\omega '). \label{eq:P0}
\ee
We have replaced the upper limit of the original integral $\infty$
with $\omega$ because $P_0^>(\omega) = 0$ for negative $\omega$
at zero temperature.

We now solve the above equation numerically following the procedures
described below.  We partition the frequency
space into equal parts with step size $\Delta \omega \ll \Delta E$
using division points $\omega_i$.  Then the
function $P_0^>(\omega)$ is replaced by an array of
numbers $P_0^>(\omega_i)$ and
the above integral equation by a matrix equation.  Instead of
inverting a huge matrix, we may calculate $P_0^>(\omega_i)$
by solving an elementary first order algebraic equation if we
know $P_0^>(\omega_j)$ for all $\omega_j < \omega_i$.  Since we
know $P_0^>(\omega)\propto\omega$
in the low frequency limit where $\omega \ll \Delta E$,
(see Sec.~\ref{sec:pt})
we may use a linear function in a small low frequency range as a `seed' to
start sequential calculations of the rest of the whole range of
interest.  Note that we cannot determine a multiplicative overall constant
in computing $P_0^>(\omega)$ because the integral equation is homogeneous.

\end{document}